\documentclass[11pt,twoside]{article}
\usepackage{macro-fsut-eng}
\usepackage{psfig}
\usepackage{graphicx}

\usepackage[T1]{fontenc} 

\usepackage{latexsym}
\usepackage{verbatim}

\begin{document}

\vskip 1.0cm
\markboth{V.~Bromm}{First Stars and Galaxies}
\pagestyle{myheadings}

\vspace*{0.5cm}
\title{The First Stars and Galaxies -- Basic Principles}

\author{V.~Bromm}
\affil{University of Texas, Department of Astronomy, Austin, TX 78712, U.S.A.}

\begin{abstract}
Understanding the formation of the first stars and galaxies is a key
problem in modern cosmology. In these lecture notes, we will derive
some of the basic physical principles underlying this emerging field. 
We will consider the basic cosmological context, the cooling and
chemistry in primordial gas, the physics of gravitational instability,
and the main properties of the first stars. We will conclude with
a discussion of the observational signature of the first sources of light,
to be probed with future telescopes, such as the {\it James Webb
Space Telescope}.
\end{abstract}

\section{Introduction}

The first sources of light fundamentally transformed the early universe,
from the simple initial state of the cosmic dark ages into one of
ever-increasing complexity (for comprehensive reviews, see Barkana \&
Loeb 2001; Bromm et al. 2009; Loeb 2010; Whalen et al. 2010).
This process began with the formation of
the first stars, the so-called Population~III (Pop~III), at redshifts
$z\sim 20-30$. These stars are predicted to form in dark matter
minihalos, comprising total masses of $\sim 10^6 M_{\odot}$. Current 
models suggest that Pop~III stars were typically massive, or even very
massive, with $M_{\ast}\sim 10-100 M_{\odot}$; these models also
predict that the first stars formed in small groups, including
binaries or higher-order multiples. First star formation has been 
reviewed in Bromm \& Larson (2004) and Glover (2005). There, you will
find a survey of the key papers up to 2005. For the most recent
developments, I recommend to consult the specialized literature (e.g.,
Turk et al. 2009; Stacy et al. 2010; Clark et al. 2011, Greif et al. 2011).

Once the first stars had formed, feedback processes began to modify
the surrounding intergalactic medium (IGM). It is useful to classify them
into 3 categories (see Ciardi \& Ferrara 2005): radiative, mechanical, and
chemical. The first feedback consists of the hydrogen-ionizing photons
emitted by Pop~III stars, as well as the less energetic, molecule-dissociating
radiation in the Lyman-Werner (LW) bands. Once the first stars die, after
their short life of a few million years, they will explode as a supernova (SN),
or directly collapse into massive black holes (MBHs). In the SN case,
mechanical and chemical feedback come into play. The SN blastwave exerts
a direct, possibly very disruptive, feedback on its host system, whereas the
chemical feedback acts in a more indirect way, as follows: The first stars,
forming out of metal-free, primordial gas, are predicted to be characterized
by a top-heavy initial mass function (IMF). Once the gas had been enriched
to a threshold level, termed ``critical metallicity'' ($Z_{\rm crit}$),
the mode of star formation would revert to a more normal IMF, which is
dominated by lower mass stars (see Frebel et al. 2007). Chemical feedback refers to this transition
in star formation mode, implying that less massive stars have a less
disruptive impact on their surroundings. The complex physics of
pre-galactic metal enrichment, and the nucleosynthesis in Pop~III SNe, are
reviewed in Karlsson et al. (2012).

The first, Pop~III stars are thus predicted to form in small groups,
in minihalos with shallow gravitational potential wells. As they
were typically very massive, they would quickly exert a strong negative
feedback on their host systems. Numerical simulations indicate that
this feedback completely destroys the host, in the sense of heating and
evacuating all remaining gas. There would therefore be no opportunity
for a second burst of star formation in a minihalo. Furthermore, since
all (most?) Pop~III stars are massive enough to quickly die, there would
be no long-lived system of low-mass stars left behind. The Pop~III forming
minihalos, therefore, are {\it not} galaxies, if a bona-fide galaxy
is meant to imply a long-lived stellar system, embedded in a dark matter
halo. The question: {\it What is a galaxy, and, more specifically, what
is a first galaxy?}, however, is a matter of ongoing debate (see Bromm et al. 2009). And as we
have seen, this question is intimately tied up with the feedback from the
first stars, which in turn is governed by the Pop~III IMF (top-heavy or
normal). 

Theorists are currently exploring the hypothesis that ``atomic
cooling halos'' are viable hosts for the true first galaxies (Oh \&
Haiman 2002). These halos have deeper potential wells, compared to
the minihalos mentioned above; indeed, they have `virial temperatures'
of $T_{\rm vir}\simeq 10^4$\,K, enabling the primordial gas to cool
via efficient line emission from atomic hydrogen. We will further
clarify these concepts below. It is useful to keep in mind that
observers and theorists often employ different definitions. As a 
theorist, you wish to identify the first, i.e., lowest-mass, dark matter halos that 
satisfy the conditions for a galaxy. Observers, on the other hand, usually
aim at detecting truly metal-free, primordial systems. Recent simulation
results, however, suggest that such metal-free galaxies do not exist. The
reason being that rapid SN enrichment from Pop~III stars, formed
in the galaxy's minihalo progenitors, provided a bedrock of heavy elements.
Any second generation stars would then already belong to Population~II
(Pop~II). These questions, and the problem of first galaxy formation
in general, have been reviewed in Bromm \& Yoshida (2011), where the reader can
again find pointers to the detailed literature. For a comprehensive
overview of galaxy formation and evolution in general, including
the situation at lower redshifts, $0<z<5$, see the monograph by
Mo et al. (2010).

The first star and galaxy field is just entering a dynamic phase
of rapid discovery. This development is primarily driven by new 
technology, on the theory side by ever more powerful supercomputers,
reaching peta-scale machines, and on the observational side by
next-generation telescopes and facilities. Among them are the
{\it James Webb Space Telescope (JWST)}, planned for launch in
$\sim$\,2018, and the suite of extremely large, ground-based
telescopes, such as the GMT, TMT, and E-ELT. The capabilities
of the {\it JWST} are summarized in Gardner et al. (2006), as well
as in the monograph by Stiavelli (2009).
Complementary to
them are ongoing and future meter-wavelength radio arrays, designed
to detect the redshifted 21cm radiation from the neutral hydrogen
in the early universe (see the review by Furlanetto et al. 2006).
A further intriguing window into the epoch of the first stars
is provided by high-redshift gamma-ray bursts (GRBs). These are
extremely bright, relativistic explosions, triggered when a rapidly
rotating massive star is collapsing into a black hole (see the
monograph by Bloom 2011). The first stars are promising GRB
progenitors, thus possibly enabling what has been termed ``GRB
cosmology'' (for details see Kouveliotou et al. 2012).

There is a second approach to study the ancient past,
nicely complementary to the {\it in situ} observation of high-redshift
sources. This alternative channel, often termed ``Near-Field Cosmology'' (Freeman \& Bland-Hawthorn 2002), is provided by local fossils that have
survived since early cosmic times. Among them are extremely metal-poor stars
found in the halo of the Milky Way. The idea here is to scrutinize their
chemical abundance patterns and derive constraints on the properties
of the first SNe, and, indirectly, of the Pop~III progenitor stars, such
as their mass and rate of rotation (for reviews, see Beers \& Christlieb 2005;
Frebel 2010). Another class of relic objects is made up of the newly
discovered extremely faint dwarf galaxies in the Local Group. These
ultra-faint dwarf (UFD) galaxies consist of only a few hundred stars, and
reside in very low-mass dark matter halos. Their chemical and structural
history is therefore much simpler than what is encountered in massive,
mature galaxies, and it should be much more straightforward to 
make the connection with the primordial building blocks (e.g., Salvadori
\& Ferrara 2009).

The plan for these lecture notes is to provide a theoretical
`toolkit', containing the basic physical principles that
are the foundation to understand the end
of the cosmic dark ages. We will consider the overall cosmological
context, the fundamentals of star formation as applicable to
the primordial universe, the properties of the first stars, and
the physical principles underlying the assembly of the first
galaxies. We conclude with some useful tools of observational
cosmology, allowing us to connect theory with empirical probes.
For a detailed discussion of the phenomenology and the numerical simulations, we refer the reader
to the review papers and monographs cited above. Again, the goal here is to
focus on the basic framework, which will likely be relevant for many
years to come, enabling the student to follow the rapid progress unfolding
in the research literature. I invite you to join in on this grand journey
of discovery!

\section{The Cosmological Context}
\label{context}
\subsection{CDM Structure Formation}
We now have a very successful model that describes the expansion history of
the universe, and the early growth of density fluctuations (Komatsu et al. 2011). This is
the $\Lambda$ cold dark matter ($\Lambda$CDM) model, as calibrated to 
very high precision by the {\it Wilkinson Microwave Anisotropy Probe (WMAP)}.
Structure formation in this model proceeds hierarchically, in a bottom-up
fashion, such that small objects emerge first, and subsequently grow
through mergers with neighboring objects and the smooth accretion of matter.
To characterize the resulting distribution of density fluctuations, we measure
the ``overdensity'' in a spherical window of radius $R$ and total (gas $+$ dark
matter) mass $M$, where $M=4\pi/3 \bar{\rho}R^3$:
\begin{equation}
\delta_M\equiv\frac{\rho-\bar{\rho}}{\bar{\rho}}\mbox{\ .}
\end{equation}
Here, $\rho$ is the mass density within a given window, and $\bar{\rho}$ that
of the background universe at the time the overdensity is measured.
Next, the idea is to place the window at random everywhere in the universe, and
to calculate the (mean-square) average: $\sigma^2(M)=\langle \delta^2_M\rangle$,
where the brackets indicate a spatial average. The latter is closely related to the
ensemble average, where one considers many realizations of the underlying
random process that generated the density fluctuations in the very early universe
(ergodic theorem). Here and in the following, all spatial scales are physical,
as opposed to comoving, unless noted otherwise.

Due to gravity, the density perturbations grow in time. This growth is described
with a ``growth factor'', $D(z)$, such that:
\begin{equation}
\sigma \propto D(z) \propto a=\frac{1}{1+z}\mbox{\ .}
\end{equation}
The second proportionality is only approximate, and would be strictly valid
in a simple Einstein-de Sitter background model. The expression for the growth
factor is more complicated in a $\Lambda$-dominated universe (see Loeb 2010),
but the Einstein-de Sitter scaling still gives a rough idea for what is going
on at $z\gg 1$. Indeed, it is quite useful for quick back-of-the-envelope
estimates. Early on, all fluctuations are very small, with $\delta_M\ll 1$; but
at some point in time, a given overdensity will grow to order unity. One says
that a fluctuation is in its linear stage, as long as $\delta_M<1$, and becomes
``non-linear'' when $\delta_M>1$. Formally, a critical overdensity of
$\delta_c=1.69$ is often used to characterize the transition. The behavior and
evolution of the perturbations in their linear stage can be treated analytically,
e.g., by decomposing a density field into Fourier modes. Once the fluctuation
turns non-linear, one needs to resort to numerical simulations to further
follow them to increasingly high densities.

A basic tenet of modern cosmology is that the quantum-mechanical processes
that imprinted the density fluctuations in the very early universe left behind
a (near-) Gaussian random field. The probability that an overdensity has a
given value, around a narrow range $d\delta_M$ is then:
\begin{equation}
P(\delta_M)d\delta_M=\frac{1}{\sqrt{2\pi\sigma^2_M}}\exp\left[-\frac{\delta^2_M}{2\sigma^2_M}\right]d\delta_M \mbox{\ .}
\end{equation}
One speaks of a ``$\nu$-sigma peak'', when: $\delta_M=\nu \sigma_M$. Note that high-sigma 
peaks are increasingly unlikely, and therefore rare. One also says that such peaks are
highly biased, and one can show that such peaks are strongly clustered (see, e.g.,
Mo et al. 2010). The sites for the formation of the first stars and galaxies
are predicted to correspond to such high-sigma peaks.
To predict the redshift of collapse, or ``virialization'' redshift (see below), we demand:
$\delta_M(z)=D(z)\delta_M(z=0)\simeq \delta_c$, or, using equ.~(2)
\begin{displaymath}
\frac{\nu \sigma_M(z=0)}{1+z_{\rm vir}}\simeq 1.69\mbox{\ ,}
\end{displaymath}
such that: $1+z_{\rm vir}\simeq \nu \sigma_M(z=0)/1.69$, where
$\sigma_M(z=0)$ is the rms density fluctuation, extrapolated to the present.
On the scale of a minihalo ($M\sim 10^6 M_{\odot}$), one has: $\sigma_M(z=0)\sim 10$.
For collapse (virialization) to occur at, say, $z_{\rm vir}\simeq 20$, we would then
need $\nu\simeq 3.5$. Thus, the first star forming sites are rare, but not yet
so unlikely to render them completely irrelevant for cosmic history.

\subsection{Virialization of DM Halos}

Once a given perturbation becomes non-linear ($\delta_M\sim 1$), the corresponding
dark matter (DM) collapses in on itself through a process of violent dynamical
relaxation. The rapidly changing gravitational potential, $\partial \varphi/\partial t$,
acts to scatter the DM particles, and their ordered motion is converted into random
motion. The result of this ``virialization'' is a, roughly spherical, halo, where
the kinetic and gravitational potential energies approach virial equilibrium:
$2 E_{\rm kin}\simeq - E_{\rm pot}$. Note that the total energy, $E_{\rm tot}=
E_{\rm kin}+E_{\rm pot}=- E_{\rm kin}$, is negative, which implies that the
halo is bound.

It is now convenient to define the gravitational potential (potential energy
per unit mass), as follows:
\begin{equation}
\varphi = \frac{E_{\rm pot}}{M_{\rm h}}\simeq - \frac{GM_{\rm h}}{R_{\rm vir}}\mbox{\ .}
\end{equation}
Here, $M_{\rm h}$ is the halo mass (gas $+$ DM), which is connected
to the halo density and radius, often called ``virial'' density and radius, via
\begin{equation}
M_{\rm h}\simeq \frac{4\pi}{3}\rho_{\rm vir}R_{\rm vir}^3  \mbox{\ .}
\end{equation}
The virial density, established after the virialization process is complete,
is related to the background density of the universe at the time of collapse,
at $z_{\rm vir}$: $\rho_{\rm vir}\simeq 200 \bar{\rho}(z_{\rm vir})$.
In terms of the present-day background density, one has:
\begin{equation}
\bar{\rho}(z) =(1+z)^3\bar{\rho}(z=0)=2.5\times 10^{-30}\mbox{\,g\,cm$^{-3}$\,} (1+z)^3  \mbox{\ .}
\end{equation}

A very useful concept to gauge how the baryonic (gaseous) component will behave when
falling into the DM halos mentioned above is the ``virial temperature''. The idea is
to ask what would happen to a proton, of mass $m_{\rm H}=1.67\times 10^{-24}$\,g, when
it is thrown into such a DM potential well. Through compressional heating, either
adiabatically or involving shocks, the particle would acquire a random kinetic energy
of:
\begin{equation}
k_{\rm B}T_{\rm vir}\simeq \epsilon_{\rm kin}\simeq -\epsilon_{\rm pot} 
\simeq \frac{GM_{\rm h}m_{\rm H}}{R_{\rm vir}}  \mbox{\ ,}
\end{equation}
where $k_{\rm B}$ is Boltzmann's constant. Combining the equations above yields
\begin{equation}
T_{\rm vir}\simeq 10^4\mbox{\,K\,} \left(\frac{M_{\rm h}}{10^8 M_{\odot}} \right)^{2/3}\left(\frac{1+z_{\rm vir}}{10}\right)
\mbox{\ ,}
\end{equation}
where the normalizations are appropriate for a first-galaxy sytem, or,
technically, and atomic cooling halo. For a minihalo, where $M_{\rm h}
\sim 10^6 M_{\odot}$ and $z_{\rm vir}\sim 20$, one has: $T_{\rm vir}\sim 1,000$\,K.
A related quantity is the halo binding energy
\begin{equation}
E_{\rm b}\simeq |E_{\rm tot}|\simeq \frac{1}{2}\frac{G M_{\rm h}^2}{R_{\rm vir}}\simeq
10^{53}\mbox{\,ergs\,}\left(\frac{M_{\rm h}}{10^8 M_{\odot}}\right)^{5/3}
\left(\frac{1+z_{\rm vir}}{10}\right)\mbox{\ ,}
\end{equation}
where the normalizations are again appropriate for an atomic cooling halo.
For a minihalo, the corresponding number is: $E_{\rm b}\sim 10^{50}$\,ergs.
Comparing these values with the explosion energy of Pop~III SNe, where
$E_{\rm SN}\simeq 10^{51}-10^{52}$\, ergs, one gets the zero-order
prediction that minihalos may already be severely affected by SN feedback,
evacuating most of the gas from the DM halo (see Ciardi \& Ferrara 2005). 
The more massive atomic cooling halos, on the other hand, are expected to
survive such negative SN feedback. This expectation is roughly born out
by numerical simulations (see Bromm \& Yoshida 2011).

\subsection{Gas Dissipation}

\begin{figure}  
\begin{center}
\hspace{0.1cm}
\psfig{figure=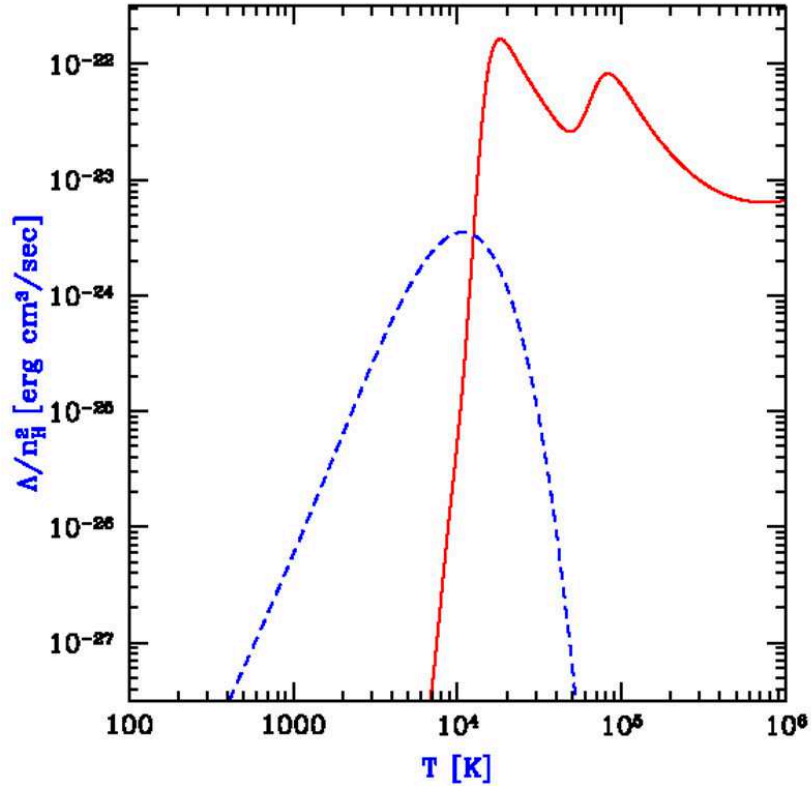,width=0.8\textwidth}
\caption{Cooling rate of primordial gas as a function of temperature.
Shown is the contribution from atomic hydrogen and helium ({\it solid line}),
as well as that from molecular hydrogen ({\it dashed line}). Atomic hydrogen 
line cooling is very efficient at temperatures of $T> 10^4$\,K, whereas
at lower temperatures, cooling has to rely on H$_{2}$, which is a poor
coolant. This is the regime of the minihalos, hosting the formation
of the first stars. Adopted from Loeb (2010).}
\end{center}
\label{cooling}
\end{figure}

To form something interesting, such as stars, black holes, or galaxies,
gas needs to be able to collapse to high densities. Initially, such
collapse is triggered by the DM potential well in halos, as the DM
is dynamically dominant, and the gas (the baryons) just follow along.
However, different from DM, the gas is collisional, and therefore subject
to compressional heating. If this heat could not be radiated away, or
dissipated, there would eventually be sufficient pressure support to
stop the collapse. The key question then is: {\it Can the gas sufficiently
cool?} A simple, but intuitively appealing and useful, answer is provided
by the cassical Rees-Ostriker-Silk criterion, that the cooling time has
to be shorter than the free-fall time: $t_{\rm cool}<t_{\rm ff}$.
If this criterion is fulfilled, a gas cloud will be able to condense to
high densities, and possibly undergo gravitational runaway collapse.
These important timescales are defined as follows: $t_{\rm ff}\simeq
1/\sqrt{G\rho}$ and $t_{\rm cool}\simeq n k_{\rm B} T/\Lambda$, where
$\Lambda$ is the cooling function (in units of ergs\,cm$^{-3}$\,s$^{-1}$).

In Figure~1, the cooling function for primordial, pure H/He, gas is shown.
One can clearly distinguish two distinct cooling channels, one at $T>10^4$\,K,
where cooling relies on atomic hydrogen lines, and one at lower temperatures,
where the much less efficient H$_{2}$ molecule is the only available
coolant. In the present-day interstellar medium (ISM), metal species would
dominate cooling in this low-temperature regime, but, by definition, they
are absent in the primordial universe. The first cooling channel governs
the formation of the first galaxies (atomic cooling halos), since $T_{\rm vir}
\sim 10^4$\,K for $M_{\rm h}\sim 10^8 M_{\odot}$ and $z_{\rm vir}\sim 10$.
First star formation in minihalos, on the other hand, is governed by
the low-temperature, H$_{2}$, cooling channel. The reason is again
that minihalos typically have $T_{\rm vir}\sim 1,000$\,K.

\subsection{Halo Angular Momentum}

Another important ingredient for early star and galaxy formation is
angular momentum. Current cosmological models posit that the post-recombination
universe, at $z<1,000$, was free of any circulation ($\nabla \times \vec{v}=0$).
Angular momentum is thought to have been created through tidal torques
during the collapse of slightly asymmetric overdensities. The idea is that 
neighboring perturbations exert a net torque on a given halo, thus
spinning it up. It is convenient to parameterize the resulting total
angular momentum, $J$, of a virialized halo by a ``spin parameter'':
\begin{equation}
\lambda\equiv \frac{J |E_{\rm tot}|^{1/2}}{G M_{\rm h}^{5/2}}\simeq 
\left(\frac{E_{\rm rot}}{|E_{\rm tot}|}\right)^{1/2}\mbox{\ ,}
\end{equation}
where $|E_{\rm tot}|$ is again the total halo (binding) energy, and
$E_{\rm rot}\simeq J^2/(M_{\rm h}R^2_{\rm vir})$ the total rotation energy.
Numerical simulations, studying the large-scale evolution of the DM component, 
have shown that the spin parameter is distributed in a log-normal fashion
with a mean of $\bar{\lambda}\simeq 0.04$ (see Mo et al. 2010).

For the DM component, the spin parameter is conserved during collisionless
evolution. During the dissipational collapse of the gas, however, the
spin parameter can change. In particular, the system can be driven 
towards centrifugal support, where $E_{\rm rot}\simeq |E_{\rm tot}|$.
{\it What is the radius of centrifugal support?}
Assume that the baryons collapse further, in a fixed DM halo potential.
It is then straightforward to show that the baryons have to collapse 
by a factor of $\sim \lambda^{-1}$ to reach centrifugal support: $R_{\rm cent}
\simeq \lambda R_{\rm vir}$. This then is the typical dimension of any
large-scale disk that forms inside a dark matter halo.

\section{First Star Formation: Basic Principles}
\label{formation}
Primordial star formation shares many similarities with the present-day
case, in terms of basic principles. It is therefore always a good idea
to seek guidance from the rich phenomenology and understanding reached
in classical star formation theory (see McKee \& Ostriker 2007;
Zinnecker \& Yorke 2007; Bodenheimer
2011 for reviews). In the following, in the spirit of our basic toolkit 
approach, we will address gravitational (Jeans) instability, the physics
of protostellar accretion, and the properties 
of a generic stellar IMF.

\subsection{Gravitational Instability}

Consider a gaseous cloud of linear size $L$ with a given mass density $\rho$
and temperature $T$. Such a cloud will be unstable to gravitational
runaway collapse, if: $t_{\rm sound}>t_{\rm ff}$, with $t_{\rm sound}\simeq
L/c_s$ being the sound-crossing time. The sound speed is $c_s\simeq
\sqrt{k_{\rm B}T/m_{\rm H}}\propto T^{1/2}$. The intuition here is that
the free-fall time measures the strength of gravity, in the sense that a
smaller $t_{\rm ff}$ corresponds to a stronger force of gravity. Similarly,
the sound-crossing timescale provides a measure for the strength of the
opposing thermal pressure, where again a smaller $t_{\rm sound}$ indicates
stronger pressure forces. The above timescale criterion for gravitational instability can be written as:
\begin{equation}
\frac{L}{c_s} > \frac{1}{\sqrt{G\rho}} \mbox{\ .}
\end{equation}
This inequality defines the Jeans length
\begin{equation}
L>L_{\rm J} \simeq  \frac{c_s}{\sqrt{G\rho}} \mbox{\ ,}
\end{equation}
with the interpretation that a density perturbation has to exceed
a certain critical size, such that gravitational forces take over, and
cannot be balanced by thermal pressure any longer. One then defines
the {\it Jeans mass}, as follows
\begin{equation}
M_{\rm J}\sim \rho L_{\rm J}^3 \simeq 
500 M_{\odot} \left(\frac{T}{200\mbox{\,K}}\right)^{3/2}
\left(\frac{n}{10^4\mbox{\,cm}^{-3}}\right)^{-1/2} \mbox{\ ,}
\end{equation}
where $n\simeq \rho/m_{\rm H}$ is the hydrogen number density, and
the normalizations reflect typical values in Pop~III star forming regions.
A closely related concept is the {\it Bonnor-Ebert mass}, where:
$M_{\rm J}\sim 2\times M_{\rm BE}$. 

\subsection{Protostellar Accretion}

Every star, regardless of whether we are dealing with Pop~I or 
Pop~III, is assembled in an inside-out fashion, such that
a small hydrostatic protostellar core is formed first at the
center of a Jeans-unstable cloud. This initial core subsequently
grows through accretion. Feedback effects from the growing 
protostar will eventually terminate this process, thus setting the
final mass scale of the star. The initial core mass is small, close
to the so-called {\it opacity limit for fragmentation}: $M_{F}\simeq
10^{-2} M_{\odot}$. This lower limit to the mass of a star can be derived
by considering a cloud that is roughly in free-fall collapse. Collapse
can proceed as long as the gas is able to radiate away the concomitant
compressional heat. At very high densities, however, the gas becomes
opaque to this cooling radiation. Soon thereafter thermal pressure
forces can stop the collapse, and a hydrostatic core is born.
The key question now is to address the growth via accretion, which happens
in two distinct regimes, first as predominantly spherical accretion, and
then as disk-dominated accretion.
\subsubsection{Spherical accretion}
For a roughly spherical gas cloud with a mass close to the Jeans mass,
one can estimate the average accretion rate as follows:
\begin{equation}
\dot{M}_{\rm acc}\sim \frac{M_{\rm J}}{t_{\rm ff}}\propto
\frac{T^{3/2}\rho^{-1/2}}{\rho^{-1/2}}\sim T^{3/2}\mbox {\ .}
\end{equation}
In the first step, we have assumed that gravity cannot move any material
faster than the free-fall time. Using $c_s\propto T^{1/2}$, it is
sometimes convenient to rewrite this as: $\dot{M}\simeq c_s^3/G$.
Typically, one has for Pop~I: $T\sim 10$\,K $\Rightarrow \dot{M}\sim
10^{-5} M_{\odot}\mbox{\,yr}^{-1}$; and for Pop~III:
$T\sim 300$\,K $\Rightarrow \dot{M}\sim
10^{-3} M_{\odot}\mbox{\,yr}^{-1}$. This two-order of magnitude difference
in accretion rates is the basic physical reason why the first stars are
believed to be more massive than present-day stars. Early on, protostellar
accretion is mainly spherical, but with time, material with higher specific
angular momentum ($J/m$) falls in, and results in the formation of a
protostellar disk. The bulk of the accretion is then shifted to a disk
mode.

\subsubsection{Disk accretion}
Consider a circular shell, located at radius $r$, within a disk of
surface mass density $\Sigma=M_{\rm disk}/(\pi R^2_{\rm disk})$. For
such a shell, one has
\begin{equation}
dM\simeq 2\pi r dr \Sigma \mbox{\ .}
\end{equation}
Dividing by $dt$ on both sides, yields the basic expression for the
disk accretion rate: $\dot{M}_{\rm acc}\simeq 2\pi r v_r \Sigma$, with $v_r=dr/dt$
being the radial (inflow) velocity. In a non-viscous disk, one
would have: $v_r=0$. Thus the reason that there is any infall is 
the presence of viscous shear forces, or, put differently, of friction.
Such shear forces enable the outward transport of angular momentum,
and the inward transport of mass.

Intuitively, to characterize viscous transport of linear momentum, one
can set:
\begin{equation}
r v_r \sim \lambda_{\rm mfp} \bar{v} \sim \nu \mbox{\ ,}
\end{equation}
with $\nu$ being the (kinematic) viscosity coefficient (in units of cm$^2$\,s$^{-1}$).
The idea is that viscous transport involves the displacement of a particle
(or fluid element) from a region of higher to lower momentum, e.g., a radial
excursion in a Keplerian disk, where $v_{\rm rot}\propto r^{-1/2}$. Such a displacement
involves, on average, a distance equal to the mean-free path, $\lambda_{\rm mfp}$,
and proceeds with the average particle velocity, $\bar{v}$. The latter is often
equal to the thermal velocity (or sound speed). It has long been realized
that the molecular (microscopic particle) viscosity is much too small to have
an effect on astrophysical systems, such as protostellar disks. Therefore, an
``abnormal'' source of viscosity is needed. Often, turbulent diffusion (transport) is
implicated. Due to the inherent difficulties involved in the physics of turbulent
transport, it is customary to employ the phenomenological $\alpha$-prescription, introduced
by Shakura \& Sunyaev (1973): $\nu = \alpha H_p c_s$, where $H_p$ is the pressure
scale-height, effectively a measure of the vertical disk height. The underlying intuition
is that one is dealing with subsonic turbulence (else, for supersonic turbulence,
shocks would rapidly dissipate any turbulent energy), such that: $\alpha < 1$. Note,
that within this picture, the turbulent eddy size is bounded by the height of the disk,
and the eddy turnover velocity by the sound speed. Also note that here, mesoscopic
fluid elements (small compared to the total system, but large compared to individual atoms)
have taken over the role of the individual atoms (or molecules) as fundamental
carriers of linear momentum transport. From numerical simulations, we have the further
constraint that, typically: $\alpha\simeq 10^{-2}-1$, depending on the nature of
the torques involved (gravitational, hydrodynamical, or magnetic).

We can now re-write our expression for the disk accretion rate in a form that can actually
be evaluated in terms of easily accessible quantities:
\begin{equation}
\dot{M}_{\rm acc}\simeq 2\pi \nu \Sigma \mbox{ .}
\end{equation}
The precise derivation would yield a ``$3\pi$'' instead of our factor of $2\pi$.
Let us conclude this subsection by considering some numbers, typically encountered
in Pop~III disks (e.g., Clark et al. 2011): $\Sigma\sim 10^2$\,g\,cm$^{-2}$, $H_p
\sim 100$\,AU\,$\sim 10^{15}-10^{16}$\,cm, and $c_s\sim 10^5$\,cm\,s$^{-1}$. Since
here, angular momentum transport is dominated by gravitational torques, it is
appropriate to choose $\alpha\sim 1$. We then estimate:
$\dot{M}_{\rm acc}\sim 10^{-3}-10^{-2} M_{\odot}$\,yr$^{-1}$. Accretion onto
a massive star proceeds for roughly the Kelvin-Helmholtz timescale, $t_{\rm acc}
\sim t_{\rm KH}\sim 10^5$\,yr, which in turn is the time it takes a (massive) star
to reach the hydrogen-burning main sequence. One then has as a rough upper
limit for the final mass of a Pop~III star: $M_{\ast}\sim \dot{M}_{\rm acc}
t_{\rm acc}\sim 100 M_{\odot}$. In reality, final masses will typically be
smaller, since accretion may be terminated earlier on due to the negative
radiative feedback from the growing protostar (e.g., Hosokawa et al. 2011;
Stacy et al. 2012).

\subsection{Initial Mass Function}

The stellar IMF is a complicated function of mass, but it is often convenient
to simply write it as a power law, valid for a given mass range. Specifically,
one considers the number of stars per unit mass:
\begin{equation}
\frac{dN}{dM}\propto M^{-x}\mbox{\ ,}
\end{equation}
where the present-day IMF is characterized by the famous Salpeter slope of $x= 2.35$.
To understand what the {\it typical} outcome of the star-formation process is, one
can ask: {\it Where does most of the available mass go?} Or, put differently:
What is the average stellar mass? This can be calculated as follows:
\begin{equation}
\bar{M}=\frac{\int_{M_{\rm low}}^{M_{\rm up}}M\frac{dN}{dM}dM}
{\int_{M_{\rm low}}^{M_{\rm up}}\frac{dN}{dM}dM}
= \frac{1-x}{2-x}\frac{M_{\rm up}^{2-x} -  M_{\rm low}^{2-x}}
{M_{\rm up}^{1-x} -  M_{\rm low}^{1-x}}\sim 3.8 \times M_{\rm low}
\end{equation}
where in the last relation, we have used the Salpeter value for $x$,
and $M_{\rm low}$ and $M_{\rm up}$ are the lower and upper mass limits, respectively.
In general, one can neglect all terms involving $M_{\rm up}$ above,
as long as $x>2$. 
This means that a Salpeter-like IMF is dominated by the
lower-mass limit. For Pop~I (present-day) stars,
one often takes $M_{\rm low}\simeq 0.1 M_{\odot}$
and $M_{\rm up}\simeq 100 M_{\odot}$, such that
$\bar{M}\sim 0.5 M_{\odot}$, whereas for Pop~III,
current theory postulates a characteristic (typical)
mass of $\bar{M}\sim 4 M_{\rm low}\sim \mbox{\, a few\,}\times 10 M_{\odot}$, assuming
that the Pop~III IMF were to exhibit a slope similar to Salpeter. The latter assumption
is not at all proven, and just serves as a zero-order guess.

\section{First Stars: Basic Properties}
In the following, we will address the fundamentals of the structure
and evolution of massive, Pop~III stars. Many of these relations are
strictly valid only for stars with $M_{\ast}>100 M_{\odot}$. The stellar
physics in this high-mass regime, although involving extreme energies
and temperatures, is of an appealing simplicity. Recent simulations have
revised the Pop~III mass-scale downwards to more modest values
($M_{\ast}\sim 10-50 M_{\odot}$), but the simple high-mass physics
nevertheless provides us with very useful order-of-magnitude
estimates. The discussion here is largely based on Bromm et al. (2001),
and we refer the reader to this paper, and references therein, for a
more detailed (numerical) treatment.
\begin{figure}  
\begin{center}
\hspace{0.1cm}
\psfig{figure=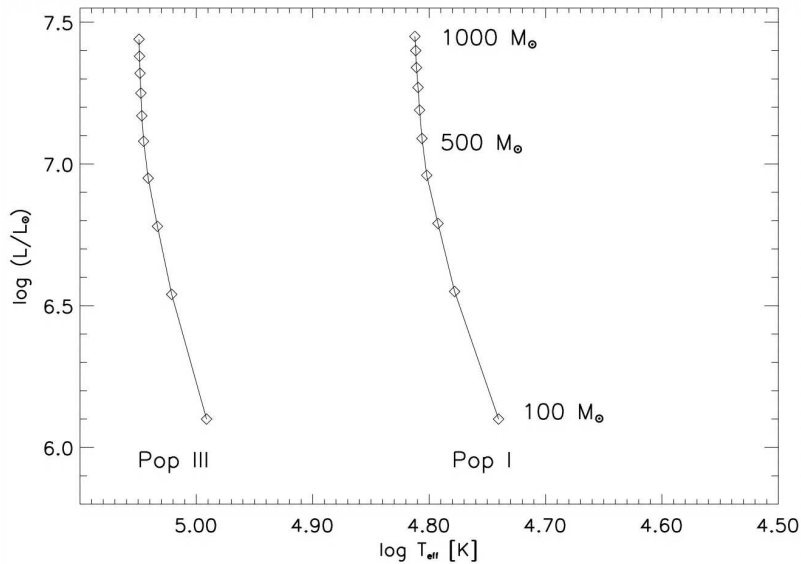,width=0.8\textwidth}
\caption{Zero-age main sequence (ZAMS) for very massive stars. Shown is a comparison
between the Pop~III ({\it left line}) and Pop~I cases ({\it right line}).
Stellar luminosity (in units of $L_{\odot}$) is plotted vs. effective temperature (in K).
{\it Diamond-shaped symbols:} Stellar masses along the sequence, from $100 M_{\odot}$ (bottom)
to $1000 M_{\odot}$ (top). As can be seen, the Pop~III ZAMS is shifted to higher values of
effective temperature, asymptotically reaching $T_{\rm eff}\simeq 10^5$\,K. Note that current
estimates for the Pop~III mass are somewhat less extreme. Adopted from Bromm et al. (2001).}
\label{HRD}
\end{center}
\end{figure}

In this spirit, we start by assuming a typical Pop~III stellar mass of
$M_{\ast}\sim 100 M_{\odot}$. Massive stars are dominated by
radiation pressure (see below), and their structure can therefore
approximately be described as a polytrope with index $n=3$. For such
a configuration, one can derive the following mass-radius relation:
\begin{equation}
R_{\ast}\simeq 5 R_{\odot} \left(\frac{M_{\ast}}{100 M_{\odot}}\right)^{1/2}
\mbox{\ ,}
\end{equation}
which is valid for a star on the hydrogen-burning main sequence.
It is then straightforward to estimate the average mass:
$\langle \rho \rangle = M_{\ast}/(4\pi/3 R^3_{\ast})\sim 1$\,g\,cm$^{-3}$,
which is of the same order as for the Sun. Somewhat surprisingly, massive stars, therefore do not
involve extreme densities. We next wish to estimate the temperature, both
at the surface and in the deep interior. To accomplish this, one has
to consider the pressure inside the star.


\subsection{Radiation Pressure}
In massive stars ($M_{\ast}>50 M_{\odot}$), radiation pressure starts
to be important, compared with the usual gas (thermal) pressure. For 
ultra-relativistic particles, in our case photons, one has the general
relation between pressure and energy density (in units of erg\,cm$^{-3}$): $P_{\rm rad}=1/3 u_{\rm rad}$.
The latter is given by the Stefan-Boltzmann law:
\begin{equation}
u_{\rm rad}=\frac{8\pi^5}{15}\frac{k_{\rm B}^4}{h^3c^3}T^4
=a_{\rm rad} T^4\mbox{\ ,}
\end{equation}
where $h$ is Planck's constant. The radiation constant has the
numerical value: $a_{\rm rad}=7.57\times 10^{-15}$\,erg\,K$^{-4}$\,cm$^{-3}$.
Thus, we can write the radiation (photon) pressure, as:
\begin{equation}
P_{\rm rad}=\frac{1}{3}
=a_{\rm rad} T^4\mbox{\ .}
\end{equation}

\subsection{Hydrostatic Equilibrium}
The motion of any fluid element in a star is governed by the Euler equation: 
\begin{equation}
\rho\frac{D\vec{v}}{Dt}=-\nabla P + \rho\vec{g}\mbox{\ ,}
\end{equation}
where $D/Dt$ is the substantial (moving with the fluid) derivative,
and $\vec{g}$ the gravitational acceleration. Assuming spherical symmetry
and $D/Dt=0$, we get the equation of hydrostatic equilibrium, which
is crucial for the structure of stars:
\begin{equation}
\frac{dP_{\rm rad}}{dr}=- \rho\frac{Gm}{r^2}\mbox{\ ,}
\end{equation}
Here, we have assumed $P=P_{\rm gas}+P_{\rm rad}\sim P_{\rm rad}$,
as is appropriate for massive stars. Above, we introduce the Lagrangian
mass coordinate, $m=m(r)$, which measures the total mass contained within
a shell of radius $r$. We now re-work this equation in an order-of-magnitude
fashion:
\begin{equation}
\frac{a_{\rm rad}T^4}{R_{\ast}}\simeq \langle\rho\rangle\frac{GM_{\ast}}{R_{\ast}^2}\mbox{\ ,}
\end{equation}
resulting in an estimate for the typical interior temperature in a massive
Pop~III star:
\begin{equation}
T_I=\left(\frac{\langle\rho\rangle}{a_{\rm rad}}\frac{GM_{\ast}}{R_{\ast}}\right)^{1/4}\sim 10^8\mbox{\,K\ .}
\end{equation}
The next question is: {\it How is this related to the surface temperature?}
In addressing it, we need to consider how energy is transported in
the stellar interior via photon diffusion.

\subsection{Radiative Diffusion}
Let us begin with a thought experiment. For the moment, (wrongly) assume
that photons are free to escape from the hot interior of a star, which
would imply negligible opacity (i.e., the capability of stellar gas to
bottle-up radiation). The timescale for such hypothetical, direct
escape would be: $t_{\rm direct}=R_{\ast}/c$. Within the same setup,
we would estimate the stellar luminosity as follows:
\begin{equation}
L=\frac{\Delta E}{\Delta t}\sim \frac{u_{\rm rad}R_{\ast}^3}{t_{\rm direct}}
\simeq \frac{a_{\rm rad}cT_I^4R_{\ast}^3}{R_{\ast}}\simeq
a_{\rm rad}cR{\ast}^2T_I^4\mbox{\ .}
\end{equation}
The radiation constant is related to the Stefan-Boltmann constant via:
$a_{\rm rad}=4\sigma_{\rm SB}/c$, where: 
$\sigma_{\rm SB}=
5.67\times 10^{-5}$\,erg\,s$^{-1}$\,cm$^{-3}$\,K$^{-4}$.
We can thus re-phrase the equation above in a form similar to the
standard blackbody expression:
\begin{equation}
L=4\pi R^2_{\ast}\sigma_{\rm SB}T_I^4\mbox{\ .}
\end{equation}

Now, why is this reasoning incorrect? The answer is that in reality 
photons are trapped inside a star. Indeed, stellar material is
typically extremely opaque to radiation, such that the photons
engage in a very slow diffusion process, and eventually leak out
from a narrow layer close to the surface, the stellar photosphere.
Effectively, we are dealing with a near-blackbody, characterized
by an effective (photospheric) temperature, $T_{\rm eff}$.
If we further replace the (incorrect) direct escape timescale with
the (correct) diffusion time, we have:
\begin{equation}
L=\frac{\Delta E}{\Delta t}\simeq \frac{\Delta E}{t_{\rm diff}}
=4\pi R^2_{\ast}\sigma_{\rm SB}T_{\rm eff}^4\mbox{\ ,}
\end{equation}
such that:
\begin{equation}
T_{\rm eff}\simeq \left(\frac{t_{\rm direct}}{t_{\rm diff}}\right)^{1/4}T_I \mbox{\ .}
\end{equation}
Our remaining task is therefore to estimate the diffusion timescale.

We can do this by modeling diffusion as a (nearly-isotropic) random
walk. If $l_{\gamma}$ is the photon mean-free path, and $N_{\rm sc}$
the number of scatterings needed for a photon to escape from the star,
basic random-walk theory yields: $R_{\ast}=\sqrt{N_{\rm sc}}l_{\gamma}$.
Stellar material is extremely opaque, and one typically has:
$l_{\gamma}\simeq 1/(n \sigma_{\rm T}) \simeq 1$\,cm. Here, we have assumed
that the opacity is dominated by electron (Thomson) scattering, with an
interaction cross-section of $\sigma_{\rm T}=0.67\times 10^{-24}$\,cm$^2$.
The diffusion timescale can then be estimated via:
\begin{equation}
t_{\rm diff}\simeq \frac{N_{\rm sc}l_{\gamma}}{c}\simeq
\frac{R_{\ast}^2}{l_{\gamma}c}\mbox{\ .}
\end{equation}
Combining everything, we finally get:
\begin{equation}
T_{\rm eff}\simeq \left(\frac{l_{\gamma}}{R_{\ast}}\right)^{1/4} T_I \sim
10^{-3}T_I\sim 10^5 \mbox{\,K\ .}
\end{equation}
Massive Pop~III stars are thus extremely hot (compare to the
solar $T_{{\rm eff},\odot}\simeq 6,000$\,K), which in turn implies
a number of key consequences for early cosmic history. Among them
is a very high specific (per unit stellar mass) production rate of
ionizing photons. It is useful to remember: $\dot{N}_{\rm ion}
\simeq 10^{48}$\,s$^{-1} M_{\odot}^{-1}$. In addition, such hot stars
can also produce copious amounts of He-ionizing photons, including
those required for the second ionization of He (54\,eV). Normally,
such extremely energetic radiation is not produced by stars, and instead
originates in quasar (accreting black-hole) sources.

\subsection{Stellar Luminosity}
Combining the expressions above, we find for the Pop~III stellar luminosity:
\begin{equation}
L=4\pi R^2_{\ast}\sigma_{\rm SB}T_{\rm eff}^4\simeq 10^6 L_{\odot}
\left(\frac{M_{\ast}}{100 M_{\odot}}\right)
\mbox{\ .}
\end{equation}
To repeat, in deriving this, we have assumed that: {\it (i)} pressure 
is dominated by radiation (appropriate for all massive stars), and {\it (ii)}
opacity is dominated by Thomson scattering (appropriate for metal-free stars).
The $L\propto M_{\ast}$ scaling is characteristic for very massive stars.
Indeed, the luminosity for massive Pop~III stars is close to the 
theoretical upper limit, the so-called ``Eddington luminosity''.

The Eddington limit can be derived by demanding that the radiation
pressure associated with a given luminosity does not exceed
the force of gravity. Specifically, let us consider the balance
of forces exerted on a combination of one proton and one electron,
where the proton provides (most of) the mass, and the (free) electron
the opacity (via Thomson scattering):
\begin{equation}
\frac{G m_{\rm H} M_{\ast}}{r^2}
\simeq \frac{\Delta p_{\gamma}}{\Delta t}\mbox{\ ,}
\end{equation}
where $\Delta p_{\gamma}$ is the absorbed photon momentum per particle:
\begin{equation}
\frac{\Delta p_{\gamma}}{\Delta t}
=\frac{\Delta E_{\gamma}/c}{\Delta t}=
\frac{1}{c}\frac{L}{4\pi r^2}\sigma_{\rm T}\mbox{\ .}
\end{equation}
Thus, one has $L<L_{\rm EDD}$, where the Eddington luminosity is:
\begin{equation}
L_{\rm EDD}=\frac{4\pi G c m_{\rm H}}{\sigma_{\rm T}}M_{\ast}=
1.3\times 10^{38}\mbox{\,erg\,s}^{-1} \left(\frac{M_{\ast}}{M_{\odot}}\right)
\mbox{\ .}
\end{equation}
This is very close to the estimate in equation~(33), demonstrating that
$L\sim L_{\rm EDD}$ for very massive ($M_{\ast}>100 M_{\odot}$) Pop~III
stars. Again, these relations still give reasonable ballpark numbers for
less extreme masses.

In Figure~2, the Pop~III zero-age main sequence (ZAMS) is shown, together
with the comparison Pop~I sequence, in both cases for very massive stars.
As is evident there, the effective temperatures asymptotically approach
constant values, $T_{\rm eff}\sim 10^5$\,K for Pop~III, and $\sim 60,000$\,K
for Pop~I. Luminosities, however, are very simlar for a star of given mass,
regardless of metallicity. 
The reason that Pop~III stars are so much hotter (bluer) is their much
more compact configuration, which is in turn a reflection of the reduced
opacity in the outer envelope.

Finally, let us estimate the lifetime of massive Pop~III stars. Assuming
that such stars radiate close to the Eddington limit, and that they are
almost fully convective, one has the simple relation:
\begin{equation}
t_{\ast}\simeq \frac{0.007 M_{\ast} c^2}{L_{\rm EDD}}\simeq 3\times 10^6\mbox{\,yr,}
\end{equation}
where the factor of $0.007$ is the efficiency of hydrogen burning.
These are very short lifetimes, compared with the solar $t_{\ast,\odot}\sim 
10^{10}$\,yr. The implication is that any feedback effects exerted on
the surrounding IGM by the Pop~III star are almost instantaneous. For
example, the pristine gas in the neighborhood of the Pop~III star
is rapidly converted into metal-enriched material, such that 
any subsequent round of star formation will already lead to Pop~II stars.
Also note that for very massive stars the stellar lifetime becomes
independent of mass.

\section{First Galaxy Assembly}
\label{assembly}
With the emergence of the first galaxies, we witness the onset of supersonic
turbulence, which is expected to have important consequences for star
formation (reviewed in Mac~Low \& Klessen 2004; McKee \& Ostriker 2007). To see this, we estimate the
Reynolds and Mach numbers, as follows. The Reynolds number measures the
relative importance of inertia and viscous forces:
\begin{equation}
Re=\frac{\mbox{inertial acceleration}}{\mbox{viscous acceleration}}\simeq
\frac{\frac{V}{T}}{\nu\frac{V}{L^2}}\simeq \frac{VL}{\nu}\mbox{\ ,}
\end{equation}
where $V$, $L$, $T=L/V$ are charcateristic velocity, length, and time scales,
respectively. For the first galaxies, we can estimate: $V\sim v_{\rm vir}\sim 10
\mbox{\,km\,s}^{-1}$, $L\sim R_{\rm vir}\sim 1$\,kpc, and $\nu\sim \lambda_{\rm mfp}
c_s\sim 10^{18}$\,cm$^2$\,s$^{-1}$. For the last estimate, we have assumed: $\lambda_{\rm mfp}
=1/(n\sigma_{\rm coll})\sim 10^{13}$\,cm, if the number density is typically $n\sim 10^3$\,cm$^{-3}$,
and we consider collisions between neutral hydrogen atoms ($\sigma_{\rm coll}\sim 10^{-16}$\,cm$^2$). For the typical particle velocity, we assume the sound-speed of H$_2$-cooled gas ($c_s\sim
1$\,km\,s$^{-1}$). The Reynolds number in the center of the first galaxies is therefore:
$Re\sim 10^9$, indicating a highly-turbulent situation. The Mach number is:
$Ma\sim V/c_s\sim v_{\rm vir}/c_s\sim 10$, indicating supersonic flows.

Supersonic turbulence generates density fluctuations in the ISM. Statistically,
these can be described with a log-normal probability density function (PDF):
\begin{equation}
f(x)dx=\frac{1}{\sqrt{2\pi \sigma^2_x}}\exp\left[-\frac{(x-\mu_x)^2}{2\sigma^2_x}
\right]dx\mbox{\ ,}
\end{equation}
where $x\equiv\ln(\rho/\bar{\rho})$, and $\mu_x$ and $\sigma^2_x$ are the
mean and dispersion of the distribution, respectively. The latter two are
connected: $\mu_x=-\sigma^2_x/2$. This relation can easily be derived
by interpreting the PDF above as a distribution (of $x$) by volume. One
then has for the volume-averaged density: $\bar{\rho}=\int \rho f(x)dx=
\bar{\rho}\int {\rm e}^x f(x)dx$, which yields the desired result. 
Numerical simulations have shown that the dispersion of the density
PDF is connected to the Mach number of the flow: $\sigma^2_x\simeq
\ln(1+0.25Ma^2)$. Inside the first galaxies, one finds values close
to $\sigma_x\simeq 1$. In Figure~3, we show an illustrative example
from a numerical simulation of isothermal, supersonic turbulence. There,
it is also evident that the self-gravity of the gas imprints a power-law
tail toward the highest densities, on top of the log-normal PDF at
lower densities, which is generated by purely hydrodynamical effects.

\section{Observational Signature}
\label{signature}
\begin{figure}  
\begin{center}
\hspace{0.1cm}
\psfig{figure=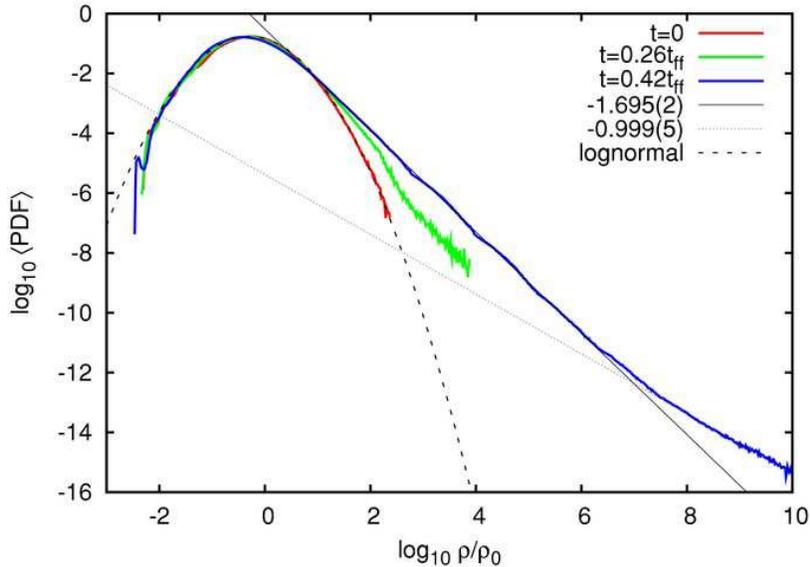,width=0.8\textwidth}
\caption{Density distribution in supersonically-turbulent flows. 
Initially ($t=0$), the density PDF is log-normal. With time, a
pronounced power-law tail develops towards high density. This is
a reflection of the gas self-gravity.
Adopted from Kritsuk et al. (2011).}
\label{TURB}
\end{center}
\end{figure}
It is useful to recall the derivations of some of the key quantities in
observational cosmology, specifically luminosity and angular-diameter distance,
as well as the observed flux. It is also useful to assemble estimates for their
typical values as encountered in the first galaxies. Further details are given
in the monographs mentioned above. In addition, a comprehensive survey of
high-redshift galaxy observations, including a description of the key
methods and tools, is given by Appenzeller (2009).

\subsection{Cosmological Distances}
In analogy to the usual inverse-square law, the luminosity distance, $d_L$, is defined via:
\begin{equation}
f_{\rm obs}= \frac{\Delta E_{\rm obs}}{\Delta t_{\rm obs} \Delta A_{\rm obs}}
=\frac{L_{\rm em}}{4\pi d^2_L}\mbox{\ ,}
\end{equation}
where here and in the following we refer to quantities that are measured at $z=0$ with
the subscript ``obs'', and with ``em'' to source-frame quantities (emitted at a given 
redshift $z$). Note that: $\Delta E_{\rm obs} = \Delta E_{\rm em}/(1+z)$, and
$\Delta t_{\rm obs} = \Delta t_{\rm em}(1+z)$ relate small differences in energy and
time in the two frames.
To evaluate $\Delta A_{\rm obs}$, carry out the following thought experiment: Imagine
that you could somehow `step outside' our universe, looking down at the scene from
some (higher-dimensional) bird's-eye perspective. Such a perspective, which of course
is completely inaccessible in practice, would allow you to measure distances, {\it
as they would appear today}. Or, put differently, you somehow managed to stop
the expansion of the universe, keep everything frozen at $z=0$, and go about
measuring the distance, with some appropriate measuring rod, from the observer (the
telescope) to where the source would be located today. Recall that the source
was much closer when it emitted the photon that we receive today, but has hence
receeded due to cosmic expansion. This source-observer distance, $r(z)$, is called {\it comoving distance}, or
{\it proper distance}. Although it cannot be directly measured, but instead can only be calculated by assuming
a theoretical model of the universe, this concept is nevertheless extremely useful in cosmology.
Assuming a point source at a given redshift $z$, we can now write the
(proper) size of the spherical surface over which the photons have been spread, as:
$\Delta A_{\rm obs}=4\pi r^2(z)$.

To calculate the comoving distance, consider the Robertson-Walker (RW) metric:
\begin{equation}
ds^2=c^2dt^2-a^2dr^2\mbox{\ ,}
\end{equation}
where we have assumed a spatially flat universe, and where $ds$ is the (invariant)
space-time interval. Since they travel along null geodesics ($ds=0$), one has for
photons: $dr=cdt/a$ (recall $a=1/(1+z)$). The RW metric describes any homogeneous,
isotropic, and expanding universe. To fully specify the background cosmological
model, we also need the Friedmann equation, governing $\dot{a}$. The latter can
in turn be derived from the Einstein field equations of general relativity (see
Mo et al. 2010), yielding:
\begin{equation}
\left(\frac{\dot{a}}{a}\right) =H(z)= H_0 \sqrt{\Omega_m (1+z)^3+\Omega_{\Lambda}}\mbox{\ ,}
\end{equation}
where $H_0$, $\Omega_m$, and $\Omega_{\Lambda}$ are the Hubble constant, the density parameter
for matter, and that for dark energy, respectively, as measured to very high precision by {\it WMAP} (Komatsu et al. 2011).
In this expression, we again assume a spatially flat universe (zero curvature).
We can now carry out the integration along the photon-geodesic:
\begin{equation}
r(z)=c \int_{0}^{z}(1+z')\left|\frac{dt}{dz'}\right|dz'= 
\frac{c}{H_0} \int_{0}^{z}\frac{dz'}{\sqrt{\Omega_m(1+z')^3+\Omega_{\Lambda}}}  \mbox{\ ,}
\end{equation}
where we use the Friedmann equation in the last step.

We have now all the ingredients in hand to find an expression for the luminosity distance:
\begin{equation}
f_{\rm obs}=\frac{1}{4\pi(1+z)^2r^2} \frac{\Delta E_{\rm em}}{\Delta t_{\rm em}}=
\frac{L_{\rm em}}{4\pi r^2 (1+z)^2}\mbox{\ .}
\end{equation}
Comparing with the definition above, we finally have: $d_L(z)=(1+z) r(z)$. It is useful
to memorize the ballpark number for a source at $z\simeq 10$, appropriate for the first galaxies:
 $d_L(z) \sim 10^2$\,Gpc. 

Next, let us derive the analogous expression for {\it angular-diameter distance}, $d_A$, where
we again start with a definition that follows basic, geometrical intuition. If a source
at $z$, having a true (proper) transverse size of $D$, is observed to have an apparent 
angular size of $\Theta$, we define: $\Theta=D/d_A$. And let us again assume our
bird's-eye perspective, as before. How would the source appear at the present-day ($z=0$),
if it had just been coasting along with the expanding universe since the time that the
photons, reaching us now, were originally emitted? The situation can be described with
a virtual triangle, where: 
\begin{equation}
\Theta=\frac{D(1+z)}{r(z)}=\frac{D}{d_A}\mbox{\ ,}
\end{equation}
giving us our result: $d_A=r(z)/(1+z)$. Note that the two fundamental distances of
observational cosmology are connected: $d_L=d_A (1+z)^2$, such that it suffices to
remember only one. For a first galaxy, where $D\sim 1$\,kpc, one finds: $\Theta\sim
D/d_A\sim 1\mbox{\,kpc}/1\mbox{\,Gpc}\sim 10^{-6}\sim 0.2''$. The near-IR camera on-board
the {\it JWST} (NIRCam) should thus be able to marginally resolve these sources.

\subsection{Observed Fluxes}
To estimate how bright a first galaxy is likely to be, we need to consider
the observed specific flux (flux per unit frequency):
\begin{equation}
f_{\nu,{\rm obs}}= \frac{\Delta E_{\rm obs}}{\Delta t_{\rm obs} \Delta A_{\rm obs}
\Delta \nu_{\rm obs}}=
\frac{\Delta E_{\rm em}/(1+z)}{4\pi r^2\Delta t_{\rm em}
\Delta \nu_{\rm em}}\simeq(1+z)\frac{L_{\nu,{\rm em}}}{4\pi d^2_L}
\mbox{\ .}
\end{equation}
To arrive at a zeroth-order guess, we assume that the total stellar mass involved
in the starburst at the center of a first galaxy is: $M_{\ast}\sim 10^5 M_{\odot}$.
If we further assume that we are dealing with a top-heavy Pop~III burst, the 
stellar radiation will be characterized by $T_{\rm eff}\sim 10^5$\,K, corresponding
to a peak frquency of $\nu_{\rm max}\sim 10^{16}$\,Hz, and a total luminosity close
to the Eddington-luminosity (see Section~4): $L\sim L_{\rm EDD}\sim 10^{43}$\,erg\,s$^{-1}$.
The emitted specific luminosity is thus: $L_{\nu,{\rm em}}\sim L_{\rm EDD}/\nu_{\rm max}\sim
10^{27}$\,erg\,s$^{-1}$\,Hz$^{-1}$.
The observed (specific) flux for a first galaxy at $z\sim 10$ is then:
\begin{equation}
f_{\nu,{\rm obs}}\sim 10^{-32}\mbox{\,erg\,s}^{-1}\mbox{\,cm}^{-2}\mbox{\,Hz}^{-1}=
1\mbox{\,nJy\ .}
\end{equation}
The nJy is indeed the typical brightness level that the {\it JWST} is designed
to image with NIRCam, thus reiterating the point that with this next-generation
facility, we will get the first galaxies within reach of deep-field exposures.

\section{Outlook}
\label{outlook}
The next decade will be very exciting as we are opening up multiple
windows into the cosmic dark ages. We will finally be able to close
the remaining gap in the long quest to reconstruct the entire history
of the universe, which began with the pioneers of cosmology in the
1920s. There will be many opportunities to make important discoveries, e.g.,
in direcly detecting the first sources of light, and in working out
a well-tested theorectical framework for star and galaxy formation at 
the dawn of time. Very likely, serendipity will play a crucial role.
It is thus a good idea to equip oneself with a comprehensive set of
tools, such as the basic physics covered in these lecture notes. After all,
there is wisdom to the old adage that ``fortune favors the prepared mind''.

\acknowledgments I would like to thank the organizers for their warm
hospitality, and for putting together a stimulating program. VB acknowledges
support from NSF grant AST-1009928 and NASA ATFP grant NNX09AJ33G.


\begin{references}

\reference Appenzeller, I. 2009, High-Redshift Galaxies (Berlin: Springer)
\reference Barkana, R., Loeb, A. 2001, Physics Reports, 349, 125
\reference Beers, T.~C., Christlieb, N. 2005, \araa, 43, 531
\reference Bloom, J.~S. 2011, What are Gamma-ray Bursts? (Princeton: Princeton Univ. Press)
\reference Bodenheimer, P.~H. 2011, Principles of Star Formation (Berlin: Springer)
\reference Bromm, V., Kudritzki, R.~P., Loeb, A. 2001, \apj, 552, 464
\reference Bromm, V., Larson, R.~B. 2004, \araa, 42, 79
\reference Bromm, V., Yoshida, N., Hernquist, L., McKee, C.~F. 2009, \nat, 459, 49
\reference Bromm, V., Yoshida, N. 2011, \araa, 49, 373
\reference Ciardi, B., Ferrara, A. 2005, Space Science Reviews, 116, 625
\reference Clark, P.~C., Glover, S.~C.~O., Smith, R.~J., et~al. 2011, Science, 331, 1040
\reference Frebel, A., Johnson, J.~L., Bromm, V. 2007, \mnras, 380, L40
\reference Frebel, A. 2010, Astronomische Nachrichten, 331, 474
\reference Freeman, K., Bland-Hawthorn, J. 2002, \araa, 40, 487
\reference Furlanetto, S.~R., Oh, S.~P., Briggs, F.~H. 2006, Physics Reports, 433, 181
\reference Gardner, J.~P., Mather, J.~C., Clampin, M., et~al. 2006, Space Science Reviews, 123, 485
\reference Glover, S.~C.~O. 2005, Space Science Reviews, 117, 445
\reference Greif, T.~H., Springel, V., White, S.~D.~M., et~al. 2011, \apj, 737, 75
\reference Hosokawa, T., Omukai, K., Yoshida, N., Yorke, H.~W. 2011, Science, 334, 1250
\reference Karlsson, T., Bromm, V., Bland-Hawthorn, J. 2012, Rev. Mod. Phys., submitted (arXiv:1101.4024)
\reference Komatsu, E., Smith, K.~M., Dunkley, J., et~al. 2011, ApJS, 192, 18
\reference Kouveliotou, C., Woosley, S.~E., Wijers, R.~A.~M.~J. (eds.) 2012,
Gamma-ray Bursts (Cambridge: Cambridge Univ. Press)
\reference Kritsuk, A.~G., Norman, M.~L., Wagner, R. 2011, \apj, 727, L20
\reference Loeb, A. 2010, How did the First Stars and Galaxies Form? (Princeton: Princeton Univ. Press)
\reference Mac~Low, M.~M., Klessen, R.~S. 2004, Rev. Mod. Phys., 76, 125
\reference McKee, C.~F., Ostriker, E.~C. 2007, \araa, 45, 565
\reference Mo, H., Van~den~Bosch, F., White, S.~D.~M. 2010, Galaxy Formation and Evolution (Cambridge: Cambridge Univ. Press)
\reference Oh, S.~P., Haiman, Z. 2002, \apj, 569, 558
\reference Salvadori, S., Ferrara, A. 2009, \mnras, 395, L6
\reference Shakura, N.~I., Sunyaev, R.~A. 1973, \aap, 24, 337
\reference Stacy, A., Greif, T.~H., Bromm, V. 2010, \mnras, 403, 45
\reference Stacy, A., Greif, T.~H., Bromm, V. 2012, \mnras, in press (arXiv:1109.3147)
\reference Stiavelli, M. 2009, From First Light to Reionization: The End of the Dark Ages (Weinheim: Wiley-VCH)
\reference Turk, M.~J., Abel, T., O'Shea, B. 2009, Science, 325, 601
\reference Whalen, D.~J., Bromm, V., Yoshida, N. (eds.) 2010, The First Stars 
and Galaxies: Challenges for the Next Decade (AIP Conference Series, vol. 1294)
\reference Zinnecker, H., Yorke, H.~W. 2007, \araa, 45, 481

\end{references}
\end{document}